\documentclass[10pt,conference,a4paper]{IEEEtran}
\usepackage{times,amsmath,epsfig}
\title{Blind Adaptive Beamforming Based on Constrained Constant Modulus RLS Algorithm for Smart Antennas}
\author{%
{Lei Wang and Rodrigo C. de Lamare}%
\vspace{1.6mm}\\
\fontsize{10}{10}\selectfont\itshape
Communication Research Group, Department of Electronics, The University of York\\ York, YO10 5DD, UK\\
\fontsize{9}{9}\selectfont\ttfamily\upshape
lw517@york.ac.uk\\
rcdl500@ohm.york.ac.uk%
}
\begin{document}
\maketitle
\begin{abstract}
In this paper, we study the performance of blind adaptive beamforming algorithms for smart antennas in realistic environments. A
constrained constant modulus (CCM) design criterion is described and used for deriving a recursive least squares (RLS) type
optimization algorithm. Furthermore, two kinds of scenarios are considered in the paper for analyzing its performance.
Simulations are performed to compare the performance of the proposed method to other well-known methods for blind adaptive
beamforming. Results indicate that the proposed method has a significant faster convergence rate, better robustness to
changeable environments and better tracking capability.
\end{abstract}

\section{Introduction}
Recently, adaptive beamforming has attracted considerable attentions and been widely used in wireless communications, radar,
sonar, medical imaging and other areas \cite{Sergiy}. Many existing methods have been presented in different communication
systems \cite{Gershman}-\cite{Jian}. Blind adaptive beamforming, which is intended to form the array direction response without
knowing users' information beforehand, is an encouraging topic that deals with interference cancellation, tracking improvement
and complexity reduction.

The linearly constrained minimum variance (LCMV) method, with multiple linear constraints \cite{Trees}, is a common approach to
minimize the beamformer output power while keeping the signal of interest (SOI) from a given direction. However, because of the
required input data covariance matrix, the LCMV beamformer cannot avoid complicated computations, especially for large input
data and/or large sensor elements. Also, this method suffers from slow convergence due to the correlated nature of the input
signal \cite{Werner}.

Choi and Shim \cite{Choi} proposed another computationally efficient algorithm based on the stochastic gradient (SG) method for
finding the optimal weight vector and avoiding the estimation of the input covariance matrix. As shown in \cite{Choi}, a cost
function is optimized according to the minimum variance subject to a constraint that avoids the cancellation of the SOI, i.e.
the so-called constrained minimum variance (CMV). Nevertheless, this algorithm still cannot avoid slow convergence rate.
Furthermore, another noticeable problem is how to define the range of step size values. The small value of step size will lead
to slow convergence rate, whereas a large one will lead to high misadjustment or even instability.

In \cite{Johnson} SG algorithms using the constant modulus (CM) cost function are reviewed by Johnson \textit{et al.} for blind
parameter estimation in equalization applications. Similarly, the CM approach exploits the low modulus fluctuation exhibited by
communications signals using constant modulus constellations to extract them from the array output. Although it adapts the array
weights efficiently regardless of the array geometry, knowledge of the array manifold or the noise covariance matrix, the
CM-based beamformer is quite sensitive to the step size. In addition, the CM cost function may have local minima, and so doesn't
have closed-form solutions. Xu and Liu \cite{Xu} developed a SG algorithm based on constrained constant modulus (CCM) technique
to sort out the local minimum problem and obtain the global minima. But they still cannot find a satisfied solution in terms of
the slow convergence.

To accelerate convergence, Xu and Tsatsanis \cite{Tsatsanis} employed CMV with the recursive least squares (RLS) optimization
technique and derived the CMV-RLS algorithm. It turns out that this method exhibits improved performance and enjoys fast
convergence rate. However, it is known to experience performance degradation if some of the underlying assumptions are not
verified due to environmental effects. The signature mismatch phenomenon is one of these problems.

In this work, we propose a constrained constant modulus recursive least squares (CCM-RLS) algorithm for blind adaptive
beamforming. The scheme optimizes a cost function based on the constant modulus criterion for the array weight adaptation. We
then derive an RLS-Type algorithm that possesses better performance than those of previous methods. We carry out a comparative
analysis of existing techniques and consider two practical scenarios for assessment. Specifically, we compare the proposed
method with the existing CMV-SG \cite{Choi}, CMV-RLS \cite{Xu} and CCM-SG \cite{Miguez}.

The remaining of this paper is organized as follows. In the next section, we present a system model for smart antennas. Based on
this model, we describe the existing SG algorithms based on the constrained optimization of the minimum variance and constant
modulus cost functions in Section III. In Section IV, the RLS-type algorithms, including the proposed algorithm, are derived.
Simulation results are provided in Section V, and some conclusions are drawn in Section VI.

\section{SYSTEM MODEL}

In order to describe the system model, let us make two simplifying assumptions for the transmitter and receiver models
\cite{Dimitris}. First, the propagating signals are assumed to be produced by point sources; that is, the size of the source is
small with respect to the distance between the source and the sensors that measure the signal. Second, the sources are assumed
to be in the "far field," namely, at a large distance from the sensor array, so that the spherically propagating wave can be
reasonably approximated with a plane wave. Besides, we assume a lossless, nondispersive propagation medium, i.e., a medium that
does not attenuate the propagating signal further and the propagation speed is uniform so that the wave travels smoothly.

\begin{figure}[ht]
\begin{center}
\includegraphics[angle=0,
width=0.45\textwidth,height=0.28\textheight]{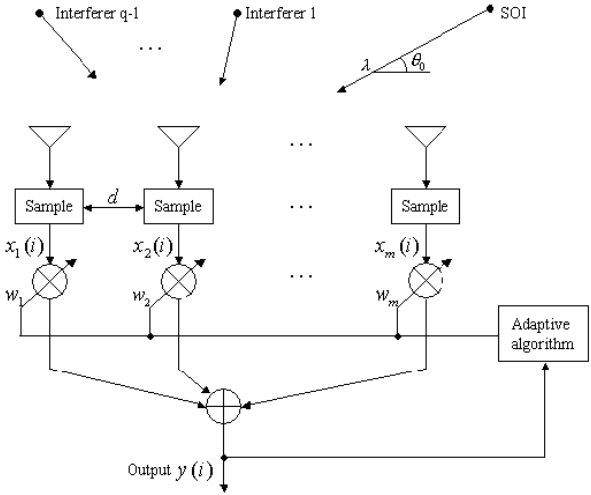}
\end{center} \caption{\label{fig:model4} Adaptive beamforming structure for ULA.}
\end{figure}

Let us consider the adaptive beamforming scheme in Fig. \ref{fig:model4} and suppose that $q$ narrowband signals impinge on the
uniform linear array (ULA) of $m$ ($q \leq m$) sensor elements from the sources with unknown directions of arrival (DOAs)
$\theta_{0}$,\ldots,$\theta_{q-1}$. The $i$th snapshot's vector of sensor array outputs can be modeled as \cite{Stoica}

\begin{equation}
\centering {\boldsymbol x}(i)={\boldsymbol A}({\boldsymbol {\theta}}){\boldsymbol s}(i)+{\boldsymbol n}(i),~~~ i=1,\ldots,N
\end{equation}
where $\boldsymbol{\theta}=[\theta_{0},\ldots,\theta_{q-1}]^{T}\in\mathcal{C}^{q \times 1}$ is the vector of the unknown signal
DOAs, ${\boldsymbol A}({\boldsymbol {\theta}})=[{\boldsymbol a}(\theta_{0}),\ldots,{\boldsymbol
a}(\theta_{q-1})]\in\mathcal{C}^{m \times q}$ is the complex matrix composed of the signal direction vectors ${\boldsymbol
a}(\theta_{i})=[1,e^{-2\pi j\frac{d}{\lambda}cos{\theta_{i}}},\ldots,e^{-2\pi
j(m-1)\frac{d}{\lambda}cos{\theta_{i}}}]^{T}\in\mathcal{C}^{m \times 1},~~~(i=0,\ldots,q-1)$, where $\lambda$ is the wavelength
and $d=\lambda/2$ is the inter-element distance of the ULA, ${\boldsymbol s}(i)\in \mathcal{R}^{q\times 1}$ is the real value
vector of the source data, ${\boldsymbol n}(i)\in\mathcal{C}^{m\times 1}$ is the complex vector of white sensor noise, which is
assumed to be a zero-mean spatially and white Gaussian process, $N$ is the number of snapshots, and $(\cdot)^{T}$\ stands for
the transpose. The output of a narrowband beamformer is given by

\begin{equation}
\centering y(i)={\boldsymbol w}^H {\boldsymbol x}(i)
\end{equation}
where ${\boldsymbol w}=[w_{1},\ldots,w_{m}]^{T}\in\mathcal{C}^{m\times 1}$ is the complex weight vector, and $(\cdot)^{H}$
stands for the Hermitian transpose.

Constrained adaptive beamformers are the ones that minimizes the objective function $J_{w}$ subject to a set of linear
constraints. That is

\begin{equation}
\centering {\boldsymbol w}=\arg\min_{w} J_{w}~~~\textrm{subject~to}~~~ {\boldsymbol C}^{H}{\boldsymbol w}={\boldsymbol f}
\end{equation}
where $\boldsymbol {w}$ as mentioned before is the vector of coefficients, ${\boldsymbol {C}}\in\mathcal{C}^{m\times q}$ is a
constraint matrix, and ${\boldsymbol f}\in\mathcal{C}^{q\times 1}$ is a constraint vector with respect to sources.

\section{Stochastic Gradient Algorithms}

The purpose of SG algorithms is to get an acceptable output performance and reduce the complexity load by avoiding the
correlation matrix estimation and inversion. We first describe the CMV and CCM algorithms on the basis of the SG method, which
are called CMV-SG and CCM-SG, respectively.  These two methods employ different cost functions, which are presented below.

\subsection{CMV-SG algorithm}

The CMV-SG algorithm chooses the beamformer to minimize the sum of the weighted array output power. The constrained optimization
means that the technique minimizes the contribution of the undesired interferences while maintaining the gain along the look
direction to be constant. For elaborating these principles clearly, we set ${\boldsymbol C}={\boldsymbol a}(\theta_{0})$, where
${\boldsymbol \theta_{0}}$ is the DOA of the desired signal, and then the received ${\boldsymbol f}$ reduces to a constant
value, say $1$. The expected result of the algorithm is to update the weight vector with respect to the input data for
minimizing the cost function of the algorithm. So, in the SG approach, the optimization problem of the CMV is represented as
follows according to the cost function

\begin{equation}
\begin{split}
\centering &J={\boldsymbol w}^{H}(i){\boldsymbol x}(i){\boldsymbol x}^{H}(i){\boldsymbol
w}(i),~~i=1,\ldots,N\\
&\textrm{subject~to}~~~{\boldsymbol w}^{H}(i){\boldsymbol a}(\theta_{0})=1
\end{split}
\end{equation}

Using the method of Lagrange multipliers, the solution can be obtained by setting the gradient term of (4) with respect to
${\boldsymbol w}(i)$ equals to zero and using the constraint ${\boldsymbol w}^{H}(i){\boldsymbol a}(\theta_{0})=1$

\begin{equation}
{\boldsymbol w}(i+1)={\boldsymbol w}(i)-\mu y^{\ast}(i)[{\boldsymbol x}(i)-{\boldsymbol a}^{H}(\theta_{0}){\boldsymbol
x}(i){\boldsymbol a}(\theta_{0})]
\end{equation}
where $\mu$ is the step size value of the CMV-SG algorithm and $\ast$ denotes complex conjugate.

\subsection{CCM-SG algorithm}

For the CCM-SG algorithm, we consider the cost function as the expected deviation of the squared modulus of the array output to
a constant, say 1. Compared with the CMV method, the CCM cost function is simply a positive measure of the average amount that
the beamformer output deviates from the unit modulus condition \cite{Jian}. By using the same constraint condition, the cost
function of CCM-SG can be expressed as

\begin{equation}
\begin{split}
\centering
&J=(|y(i)|^2-1)^2,~~i=1,\ldots,N\\
&\textrm{subject~to}~~~{\boldsymbol w}^{H}(i){\boldsymbol a}(\theta_{0})=1
\end{split}
\end{equation}

Also, using the same operation as that of CMV-SG, the update equation of weight vector can be obtained

\begin{equation}
\begin{split}
{\boldsymbol w}(i+1)&={\boldsymbol w}(i)-\mu(|y(i)|^2-1) y^{\ast}(i)\\&\quad \cdot[{\boldsymbol x}(i)-{\boldsymbol
a}^{H}(\theta_{0}){\boldsymbol x}(i){\boldsymbol a}(\theta_{0})]
\end{split}
\end{equation}

It will be shown via simulations that, by employing the constant modulus property and the constrained condition, the performance
of CCM-SG algorithm is better than that of CMV-SG one.

The SG algorithms employ stochastic inputs instead of deterministic inputs in the steepest descent method and do not use the
desired response as the least mean square (LMS) method. It is well known that the algorithms based on the SG technique generally
have low computational complexity. However, both the slow convergence behavior and the sensitivity to the step size are
drawbacks.

\section{Recursive Least Squares (RLS) Type Algorithms}

To overcome the shortcomings of the SG algorithms, we propose a fast converging RLS-type algorithm using the CCM criterion for
the array weight adaptation. An important feature of this method is that its rate of convergence is typically an order of
magnitude faster than those of the simple SG algorithms, due to the fact that RLS technique whitens the input data by using the
inverse correlation matrix of the data. Despite the initial presentation of the weight update includes the correlation matrix
inversion, recursive computations can be employed to update the value of the correlation matrix of the input data step by step.
Then, according to the matrix inversion lemma defined in \cite{Haykin}, we obtain the recursive equation for the inverse of the
correlation matrix and this reduces by one order of magnitude the high computational burden introduced by matrix inversion.

\subsection{CMV-RLS algorithm}

Here, we extend the least squares (LS) method to develop RLS-Type algorithm for solving the ill-posed problem \cite{Haykin}. In
this section, we describe the CMV-RLS algorithm first and then derive the proposed CCM-RLS method by using the same idea. The
LS-type cost function of CMV algorithm is described by

\begin{equation}
\begin{split}
\centering &J=\sum_{l=1}^{i}\alpha^{i-l}{\boldsymbol w}^{H}(i){\boldsymbol x}(l){\boldsymbol x}^{H}(l){\boldsymbol
w}(i),~~i=1,\ldots,N\\
&\textrm{subject~to}~~~{\boldsymbol w}^{H}(i){\boldsymbol a}(\theta_{0})=1
\end{split}
\end{equation}
where $\alpha$ is a positive parameter close to, but less than, unity.

Equation (8) will lead to the LS algorithm. Using the method of Lagrange multipliers, setting the gradient term of cost function
with respect to ${\boldsymbol w}(i)$, making it equal to zero and using the constraint ${\boldsymbol w}^{H}(i){\boldsymbol
a}(\theta_{0})=1$, we obtain the solution

\begin{equation}
{\boldsymbol w}(i)=[{\boldsymbol a}^{H}(\theta_{0}){\boldsymbol R}^{-1}(i){\boldsymbol a}(\theta_{0})]^{-1}{\boldsymbol
R}^{-1}(i){\boldsymbol a}(\theta_{0})
\end{equation}
where ${\boldsymbol R}(i)=\sum_{l=1}^{i}{\alpha^{i-l}}{\boldsymbol x}(l){\boldsymbol x}^{H}(l)\in\mathcal{C}^{m\times m}$.

Following the recursion for updating the correlation matrix, it can be performed as

\begin{equation}
{\boldsymbol R}(i)=\alpha{\boldsymbol R}(i-1)+{\boldsymbol x}(i){\boldsymbol x}^{H}(i)
\end{equation}

Then, as mentioned before, ${\boldsymbol R}^{-1}(i)$ can be estimated by exploiting a relation in matrix algebra as the matrix
inversion lemma. If we define ${{\boldsymbol P}(i)}={{\boldsymbol R}^{-1}(i)}$, the RLS solution of CMV-RLS is available.

\begin{equation}
{\boldsymbol w}(i)=[{\boldsymbol a}^{H}(\theta_{0}){\boldsymbol P}(i){\boldsymbol a}(\theta_{0})]^{-1}{\boldsymbol
P}(i){\boldsymbol a}(\theta_{0})
\end{equation}

Both Xu and Tsatsanis have proved that the CMV-RLS method, compared with SG algorithms, exhibits better effects on both
performance and convergence behavior in \cite{Tsatsanis}. However, these properties are adversely affected by dynamic
environments in practice, which further complicated the estimate of the statistics.

\subsection{Proposed CCM-RLS algorithm}

In order to conquer the problem occurred in CMV-RLS, we propose an improved algorithm which uses the constant modulus property.

Consider the optimization equation shown below

\begin{equation}
\begin{split}
\centering &J=\sum_{l=1}^{i}\alpha^{i-l}[|{\boldsymbol w}^{H}(i){\boldsymbol
x}(l)|^2-1]^2,~~i=1,\ldots,N\\
&\textrm{subject~to}~~~{\boldsymbol w}^{H}(i){\boldsymbol a}(\theta_{0})=1
\end{split}
\end{equation}
where $\alpha$ is similar to that in the CMV-RLS algorithm. Here, we do not need to bother with the sensitivity of step size
value since the cost function above even does not contain this term. Also, it is easy to choose $\alpha$ due to its small range.
As can be seen from the CCM-RLS cost function, a closed-form solution (global minimum) is not possible because it is a
fourth-order function with a more complicated structure, i.e. multiple local minima. The constraint condition used in (12)
ensures that we can get the optimal solution.

Equation (12) is related to the LS algorithms. Using the same operation as that of CMV-RLS, we can get the conventional solution
of our proposed method

\begin{equation}
{\boldsymbol w}(i)=[{\boldsymbol a}^{H}(\theta_{0}){\boldsymbol R}^{-1}(i){\boldsymbol a}(\theta_{0})]^{-1}{\boldsymbol
R}^{-1}(i){\boldsymbol a}(\theta_{0})
\end{equation}
where ${\boldsymbol R}(i)=2\sum_{l=1}^{i}{\alpha^{i-l}}[|{\boldsymbol w}^{H}(i){\boldsymbol x}(l)|^2-1]{\boldsymbol
x}(l){\boldsymbol x}^{H}(l)\in\mathcal{C}^{m\times m}$. LS estimation, like the method of least squares, is an ill-posed inverse
problem. To make it "well posed," we need to renew the correlation matrix ${\boldsymbol R}(i)$ \cite{Haykin} as follows

\begin{equation}
{\boldsymbol R}(i)=2\sum_{l=1}^{i}{\alpha^{i-l}}[|{\boldsymbol w}^{H}(i){\boldsymbol x}(l)|^2-1]{\boldsymbol x}(l){\boldsymbol
x}^{H}(l)+\delta\alpha^{i}{\boldsymbol I}.
\end{equation}
where $\delta$ is a positive real number called the regularization parameter and ${\boldsymbol I}$ is the $m\times m$ identity
matrix. The second term on the right side of (14) is included to stabilize the solution to the algorithm by smoothing the
solution and has the effect of making the correlation matrix ${\boldsymbol R}(i)$ nonsingular at all stages of the computation.

Let $e(i)=[|{\boldsymbol w}^{H}(i){\boldsymbol x}(l)|^2-1]$ and follow the recursion for updating the correlation matrix, (14)
can be expressed as

\begin{equation}
{\boldsymbol R}(i)=\alpha{\boldsymbol R}(i-1)+2e(i){\boldsymbol x}(i){\boldsymbol x}^{H}(i)
\end{equation}

By using the matrix inversion lemma in (15), we can obtain the inverse of ${\boldsymbol R}(i)$

\begin{equation}
\begin{split}
{\boldsymbol R}^{-1}(i)&=\alpha^{-1}{\boldsymbol R}^{-1}(i-1)\\&\quad -\frac{\alpha^{-2}{\boldsymbol R}^{-1}(i-1){\boldsymbol
x}(i){\boldsymbol x}^{H}(i){\boldsymbol R}^{-1}(i-1)}{1/2e(i)+\alpha^{-1}{\boldsymbol x}^{H}(i){\boldsymbol
R}^{-1}(i-1){\boldsymbol x}(i)}
\end{split}
\end{equation}

Here, we still define ${\boldsymbol P}(i)={\boldsymbol R}^{-1}(i)$. For convenience of computation, we also introduce a vector
${\boldsymbol k}(i)\in\mathcal{C}^{m\times 1}$ as

\begin{equation}
{\boldsymbol k}(i)=\frac{\alpha^{-1}{\boldsymbol P}(i-1){\boldsymbol x}(i)}{1/2e(i)+\alpha^{-1}{\boldsymbol
x}^{H}(i){\boldsymbol P}(i-1){\boldsymbol x}(i)}
\end{equation}

Using these definitions, we may write (16) as

\begin{equation}
{\boldsymbol P}(i)=\alpha^{-1}{\boldsymbol P}(i-1)- \alpha^{-1}{\boldsymbol
     k}(i){\boldsymbol x}^{H}(i){\boldsymbol P}(i-1)
\end{equation}

Until now, we develop a recursive equation to update the matrix ${\boldsymbol P}(i)$ by incrementing its old value. Finally,
using the fact that ${\boldsymbol P}(i)$ equals to ${\boldsymbol R}^{-1}(i)$, we get the proposed RLS solution

\begin{equation}
{\boldsymbol w}(i)=[{\boldsymbol a}^{H}(\theta_{0}){\boldsymbol P}(i){\boldsymbol a}(\theta_{0})]^{-1}{\boldsymbol
P}(i){\boldsymbol a}(\theta_{0})
\end{equation}

Equations (17)-(19), collectively and in that order, constitute the derived CCM-RLS algorithm, as summarized in Table
\ref{tab:CCM-RLS}.

\begin{table}[!t]
\centering

    \caption{CCM-RLS ALGORITHM}     
    \label{tab:CCM-RLS}

    \begin{tabular}{|l|l|}
    \hline
    Initialization & ${\boldsymbol w}(0)={\boldsymbol 0}$\\
     & ${\boldsymbol P}(0)=\delta^{-1}{\boldsymbol I}_{m\times m}$\\
     & $\delta=\left\{\begin{array}{l}
     \textrm{small positive constant for high SNR}\\
     \textrm{large positive constant for low SNR}\\
     \end{array} \right.$\\
    \hline
    Update  &${\boldsymbol \pi}(i)={\boldsymbol P}(i-1){\boldsymbol x}(i)$ \\
     & $e(i)=[|{\boldsymbol w}^{H}(i){\boldsymbol
     x}(i)|^2-1]$\\
     {(For each } & ${\boldsymbol k}(i)=\frac{{\boldsymbol \pi}(i)}{\alpha/2e(i)+{\boldsymbol x}^{H}(i){\boldsymbol
      \pi}(i)}$\\
     {instant of time} & ${\boldsymbol P}(i)=\alpha^{-1}{\boldsymbol P}(i-1)- \alpha^{-1}{\boldsymbol
     k}(i){\boldsymbol x}^{H}(i){\boldsymbol P}(i-1)$\\
      {$i=1,\ldots,N$)} & ${\boldsymbol w}(i)=[{\boldsymbol
a}^{H}(\theta_{0}){\boldsymbol P}(i){\boldsymbol a}(\theta_{0})]^{-1}{\boldsymbol P}(i){\boldsymbol
a}(\theta_{0})$ \\
    \hline
    \end{tabular}
\end{table}

We note that, (17) and (18) enable us to update the value of the vector ${\boldsymbol k}(i)$ itself. An important feature of
this algorithm described by these equations is that the inversion of the correlation matrix ${\boldsymbol R}(i)$ is replaced at
each step by a simple scalar division. Also, in the summary presented in Table \ref{tab:CCM-RLS}, the calculation of the vector
${\boldsymbol k}(i)$ proceeds in two stages:

\begin{itemize}
\item   First, an intermediate quantity, denoted by ${\boldsymbol \pi}(i)$, is computed.
\item   Second, ${\boldsymbol \pi}(i)$ is used to compute ${\boldsymbol k}(i)$.
\end{itemize}

This two-stage computation of ${\boldsymbol k}(i)$ is preferred over the direct computation of it using (16) from a
finite-precision arithmetic point of view \cite{Haykin}.

To initialize the CCM-RLS method, we need to specify two quantities:

\begin{itemize}
\item   The initial weight vector ${\boldsymbol w}(0)$. The customary practice is to set ${\boldsymbol w}(0)={\boldsymbol 0}$.
\item   The initial correlation matrix ${\boldsymbol R}(0)$. Setting $i=0$ in (14), we find that ${\boldsymbol R}(0)=\delta{\boldsymbol
I}$.
\end{itemize}

In terms of complexity, the LS requires $\mathcal {O}$($m^{3}$) arithmetic operations, whereas the RLS requires $\mathcal
{O}$($m^{2}$). Furthermore, we can notice that the step size $\mu$ in the SG algorithms is replaced by ${\boldsymbol
R}^{-1}(i)$. This modification has a significant impact on improving the convergence behavior.

\section{Simulations}

The performance of the proposed CCM-RLS algorithm is compared with three existing algorithms, namely CMV-SG, CCM-SG and CMV-RLS,
in terms of output signal-to-interference-plus-noise ratio (SINR), which is defined as

\begin {equation}
\centering {\rm SINR}(i)=\frac{{\boldsymbol w}^{H}(i){\boldsymbol R}_{s}(i){\boldsymbol w}(i)}{{\boldsymbol
w}^{H}(i){\boldsymbol R}_{i+n}(i){\boldsymbol w}(i)}
\end{equation}
where ${\boldsymbol R}_{s}(i)$ is the autocorrelation matrix of the desired signal and ${\boldsymbol R}_{i+n}(i)$ is the
cross-correlation matrix of the interference and noise in the environment.

An ULA containing $m=16$ sensor elements with half-wavelength spacing is considered. The noise is spatially and temporally white
Gaussian noise with power $\sigma_{n}^{2}=0.01$. For each scenario, $K=100$ iterations are used to get each simulated point. In
all simulations, the DOA of the SOI is $\theta_{0}=20^{o}$ and the desired signal power is $\sigma_{0}^{2}=1$. The
interference-to-noise ratio (INR), in all examples, is equal to $10 {\rm dB}$. The BPSK scheme is employed to modulate the
signals. The value of $\alpha$ was set equal to $0.998$ in order to optimize the performance of the RLS-type algorithms.

\begin{figure}[!htb]
\begin{center}
\def\epsfsize#1#2{1.02\columnwidth}
\epsfbox{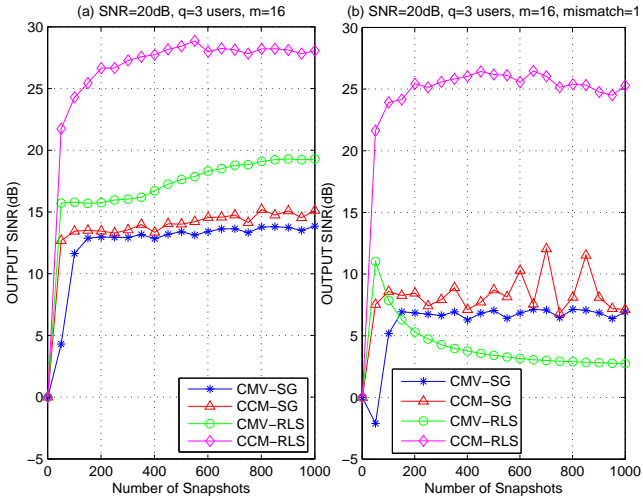} \caption{Output SINR versus number of snapshots
for (a) ideal steering vector condition (b) steering vector with
mismatch.} \label{fig:cmvccmsvefinal}
\end{center}
\end{figure}

Fig. \ref{fig:cmvccmsvefinal} includes two experiments. Fig. \ref{fig:cmvccmsvefinal}(a) shows the output SINR of each method
versus the number of snapshots, whose total is $1000$ samples. There are two interferers with DOAs $\theta_{1}=40^{o}$ and
$\theta_{2}=60^{o}$. In this environment, the actual spatial signature of the signal is known exactly. The result shows that the
RLS-type algorithms converge faster and have better performances than those of the SG algorithms. The steering vector mismatch
scenario is shown in Fig. \ref{fig:cmvccmsvefinal}(b). We assume that this steering vector error problem is caused by look
direction mismatch \cite{Sergiy}. The assumed DOA of the SOI is a random value located around the actual direction, whose
mismatch is limited in a range of $1^{o}$. Compared with Fig. \ref{fig:cmvccmsvefinal}(a), Fig. \ref{fig:cmvccmsvefinal}(b)
indicates that the mismatch problem leads to a worse performance for all the solutions. The CMV-RLS method is more sensitive to
this environment, whereas the proposed CCM-RLS algorithm is more robust to this mismatch.

\begin{figure}[!htb]
\begin{center}
\def\epsfsize#1#2{1.02\columnwidth}
\epsfbox{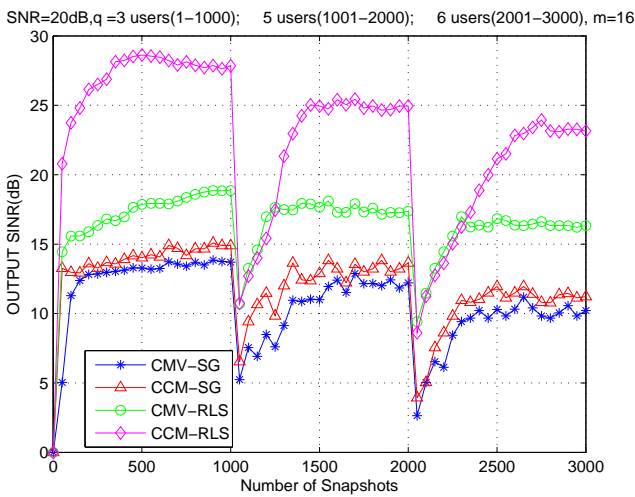} \caption{Output SINR in a scenario where
additional interferers suddenly enter and/or leave the system.}
\label{fig:moreusers3final}
\end{center}
\end{figure}

In Fig. \ref{fig:moreusers3final}, the scenario is the same as that in Fig. \ref{fig:cmvccmsvefinal}(a) for the first $1000$
samples. Two more users, whose DOAs are $\theta_{3}=30^{o}$ and $\theta_{4}=50^{o}$, enter the system in the second $1000$
samples. As can be seen from the figure, SINRs of both the SG and RLS-Type algorithms reduce at the same time. It is clear that
the performance degradation of the proposed CCM-RLS is much less significant than those of the other methods. In addition,
RLS-type methods can quickly track the change and recover to a steady-state. At $2000$ samples, two interferers with DOAs
$\theta_{5}=25^{o}$ and $\theta_{6}=35^{o}$ enter the system whereas one interferer with DOA $\theta_{4}=50^{o}$ leaves it. The
simulation shows nearly the same performance as that of the second stage. It is evident that the output SINR of our proposed
algorithm is superior to the existing techniques. This figure illustrates that the CCM-RLS is still better after an abrupt
change, in a non-stationary environment where the number of users/interferers suddenly changes in the system.

\section{Conclusions}

In this paper, a new algorithm enabling blind adaptive beamforming has been presented to enhance the performance and improve the
convergence property of the previously proposed adaptive methods. Following the CCM criterion, a RLS-type optimization algorithm
is derived. In the place of step size, we employ the correlation matrix inversion instead for increasing the convergence rate.
Then, matrix inversion lemma was used to solve this inversion problem with reduced complexity. We considered different scenarios
to compare CCM-RLS algorithm with several existing algorithms. Comparative simulation experiments were conducted to investigate
the output SINR. The performance of our new method was shown to be superior to those of others, both in terms of convergence
rate and performance under sudden change in the signal environment.

%

\end{document}